\documentclass[]{spie}  %>>> use for US letter paper
\usepackage{amsmath,amsfonts,amssymb}
\usepackage{graphicx}
\usepackage{graphics}
\usepackage{epstopdf}
\usepackage[colorlinks=true, allcolors=blue]{hyperref}

\usepackage{cite}

\title{An Adaptive software defined radio design based on a standard space telecommunication radio system API}

\author[a]{Wenhao Xiong, Xin Tian, Genshe Chen}
\author[b]{Khanh Pham}
\author[c]{Erik Blasch}
\affil[a]{Intelligent Fusion Technology, Inc. Germantown, MD, 20876}
\affil[b]{Air Force Research Laboratory, Space Vehicles Directorate, Kirtland Air Force Base, NM 87117}
\affil[c]{Air Force Research Laboratory, Information Directorate, Rome, NY 13441}

\begin{document} 
\maketitle

\begin{abstract}
Software defined radio (SDR) has become a popular tool for the implementation and testing for communications performance. The advantage of the SDR approach includes: a re-configurable design, adaptive response to changing conditions, efficient development, and highly versatile implementation. In order to understand the benefits of SDR, the space telecommunication radio system (STRS) was proposed by NASA Glenn research center (GRC) along with the standard application program interface (API) structure. Each component of the system uses a well-defined API to communicate with other components. The benefit of standard API is to relax the platform limitation of each component for addition options. For example, the waveform generating process can support a field programmable gate array (FPGA), personal computer (PC), or an embedded system. As long as the API defines the requirements, the generated waveform selection will work with the complete system. In this paper, we demonstrate the design and development of adaptive SDR following the STRS and standard API protocol. We introduce step by step the SDR testbed system including the controlling graphic user interface (GUI), database, GNU radio hardware control, and universal software radio peripheral (USRP) tranceiving front end. In addition, a performance evaluation in shown on the effectiveness of the SDR approach for space telecommunication.
\end{abstract}

% Include a list of keywords after the abstract 

\section{INTRODUCTION}
\label{sec:intro}  % \label{} allows reference to this section
Software defined radio (SDR) as a realistic emulation tool has become popular in the wireless communication field. The application ranges from the audio communication, multi-input multi-output (MIMO) communication\cite{Tian2012}, and video communication under various standards. There are many versions of the SDR using various front end hardware devices and supporting software methods. Typical systems include the universal software radio peripheral (USRP) for the hardware front end device due to its capabilities and affordability. The back-end signal processing is distributed in many components which depends on the design. In general, developers define their own architecture from which, many scenarios can be emulated using the SDR system with different designs. The cognitive space communication network proposes to integrate deep learning, cognitive radios, cognitive networking, and security using SDR \cite{Chenji2016}. The MIMO system was emulated using combination of GNU radio and USRP for the  hardware and signal processing\cite{Gardellin2012}. The Digital Video Broadcasting — Terrestrial (DVB-T) standard is validated using the SDR system with specified computer and USRP(N210) with supporting algorithms \cite{Baruffa2014}. A SDR testbed is also described with structure of fractionally spaced equalizer for synchronization and mitigation \cite{Weiss2003}. An example is for optimal data transmission from SATCOM platforms \cite{Shen2014}. Other related work can also be found in the literature \cite{Liu20152,Han2016,Liu2015,Ceylan2016,Liu2014, Steward2015, Liu20153, Macedo2015, Xia2015, Liu20162, Zhang2015,Wei2014}.

The above structures and architectures consider the interface between hardware and software components as pre-defined options. The interface is defined as high speed bus between central processing unit (CPU) and field programmable gate array (FPGA) for data exchange \cite{Chenji2016}. For example with channel sensing supported by machine learning: deep learning is performed in the computer for enhanced processing capability; while channel sensing is performed in the FPGA with the input of the learning result. The two modules' communication is limited by the system operation, for example the format of the data sometimes is the bottleneck of the information exchange mechanism. The more common case is that signal processing of all the back-end computation is assumed to be in one place, typically one personal computer (PC)\cite{Drozdenko2015, Wei2016}. This is sometimes true if the architecture of the system is simple and more importantly, if the operations of transmitting or receiving can be carried out by one node. However, this is not true for the satellite communication (SATCOM). The SATCOM scenario is a complex situation with distributed hardware locations and different software protocols that need to be well organized for robust performance. We propose to use standard APIs to accommodate the needs of the emulation.

The rest of the paper is organized as follows: Section \ref{sec:system} describes the system design architecture. Section \ref{sec:transceiving} introduces the details of the transmitter and receiver structure and implementation. Section \ref{sec:calibration} demonstrates the link calibration process utilizing proposed architecture and design. Finally, Section \ref{sec:conclusion} concludes the paper.

\section{System Model}
\label{sec:system}
\begin{figure}[t]
  \centering
    \includegraphics[width=0.7\linewidth]{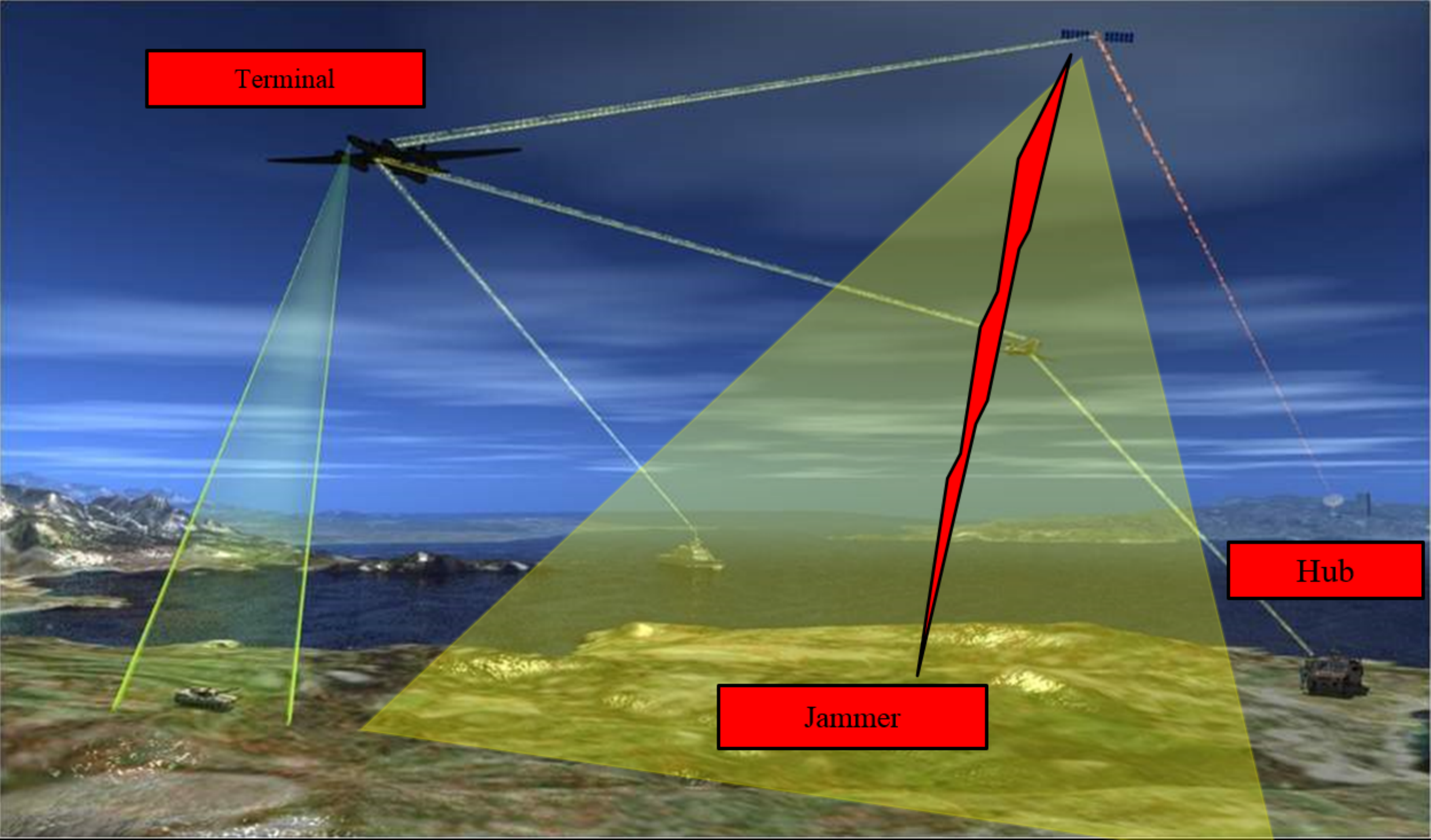}
    \caption{SATCOM scenario}
    \label{fig:scenario}
\end{figure}

Figure \ref{fig:scenario} shows an overview version of the SATCOM scenario. The \textit{terminal} establishes communication with the ground hub through a satellite transponder. The terminal is a multi-domain station from a satellite, an aircraft, to a ship. The satellite transponder could be fragile to jamming signal since many of the transponders act like a bent-pipe which only forward what is received to the destination. Hence, one major threat to SATCOM is the intentional interference signal sent out by a jammer. The jammer could be located on the ground or in the air. There are many mechanisms or algorithms proposed in the literature to deal with the jamming situation. While in this paper, we are not focused on solving the jamming problem, we are interested in emulating the complete scenario where each part of the system could be subject to jamming interference. A terminal, hub, or jammer can be modeled as an individual transmitter or receiver, and should be configured properly in order to emulate the most realistic condition of the communication. For example the carrier frequencies, bandwidth, modulation and coding type and transmitting power should be carefully defined in the emulation for each player. 

In the popular testbed design with USRP serving as the radio frequency (RF) front-end device, usually one PC controls one USRP, which is also true in our design. These PCs should be centralized organized since the scenario is usually configured with top to bottom style by a centrally device. In the meantime, the deployment of USRP is performed by each individual PC separately according to the scenario parameters. The structure requires the communication between terminals, jammers, and hubs to the scenario controlling unit, which in our case is the server. The data carried in the communication includes the payload and configuration. The payload could be binary information, I/Q symbols (i.e., quadrature signal components) and samples recorded after the filter. Configurations could be carrier frequency, bandwidth, modulation type, channel coding and others. The information is in different formation and requires different accuracy. Due to the diversity of the communication, we need broadly defined APIs between the nodes in the emulation. In this article, we provide a framework to serve all these needs.

\begin{figure}[t]
  \centering
    \includegraphics[width=0.7\linewidth]{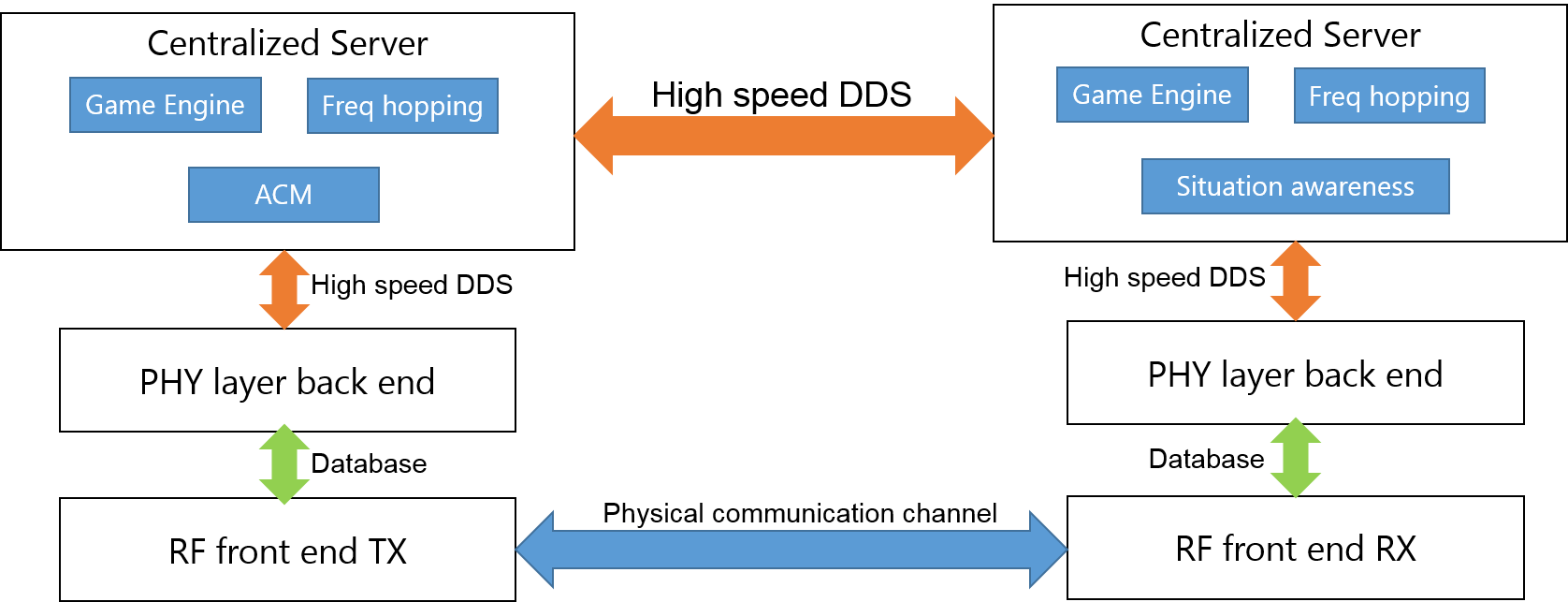}
    \caption{Emulation testbed system}
    \label{fig:system}
\end{figure}

Figure \ref{fig:system} shows the structure of our system emulation design. We separate the emulation testbed into three layers. The top layer is the server. The scenario is defined inside the server. We deploy the number of terminals, hubs, and jammers in the emulation, each with defined their transmission methods, capabilities and patterns. In our particular design, we put some of the signal processing in the server for centralized computation. For the example, in the case of a game engine where the transmitting pattern plays against the jamming pattern\cite{Wei2007, Tian20122}, frequency hopping\cite{Yu2013, Shen2012} is designed for each node; where adaptive coding and modulation uses information from the feedback channel. 

The second layer is the rest of the PHY signal processing including the channel coding, framing, modulation and pulse shaping. We are following the standard of Digital Video Broadcasting - Satellite - Second Generation (DVB-S2) which defined modulation and channel coding combinations. In our emulation, the server layer decides which combination to use in the transmission. 

The third layer is the RF front end. In this layer, hardware component USRP and antenna are configured by the software. We have control of parameters like the carrier frequency and the power amplifying ratio. These parameters are also defined by the server in order to emulate the required scenario. However, the information is passed down by the second layer through the database. It is worth noting that the power amplifying gain is controlled through a calibration tool which will be described in detail in a later section. The idea is that the server only defined the desired SNR level that is needed in a particular link, and the calibration tool will adjust the transmitting power and achieve the target SNR. Hence the amplifying gain is not directly defined by the server, but the system  will follow the instructed SNR to adjust the amplifying gain. 

In the complete emulation system, both the transmitter and the receiver side are modeled. On the server layer, they are connected with high speed data distributed service (DDS) to emulate both sides of the communication scenario. The three layers structure for the transmitter and receiver is very similar. The difference is in the server layer is from which the transmitter performs adaptive coding and modulation (ACM) and the receiver side performs situation awareness\cite{Zhang2013}.

\section{Transceiving Design}
\label{sec:transceiving}
This section presents the benefits of the standard interface in our system and introduces the details of the transceiving design.

\begin{figure}[t]
  \centering
    \includegraphics[width=0.8\linewidth]{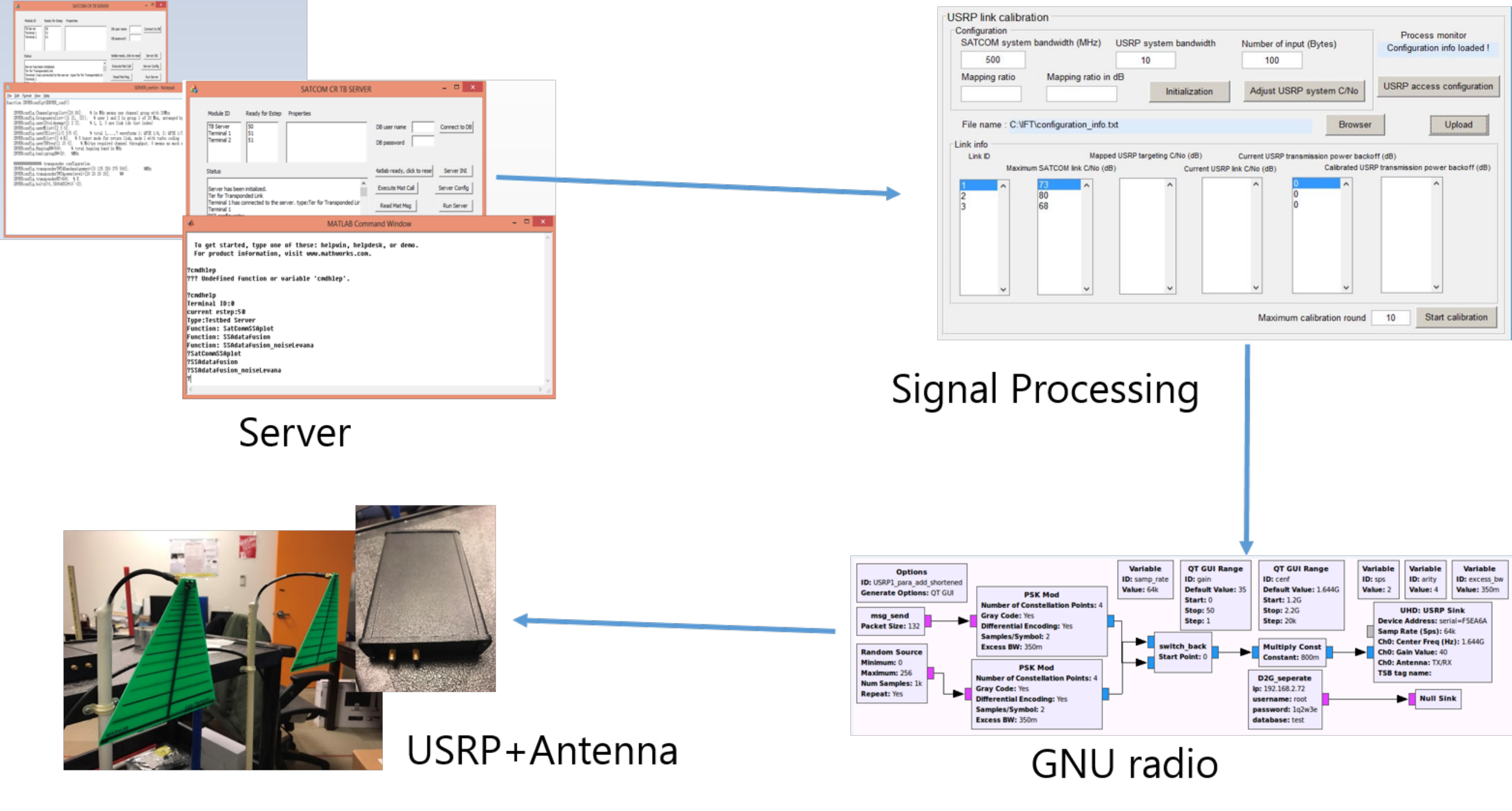}
    \caption{Transmitter}
    \label{fig:transmitter}
\end{figure}

Figure \ref{fig:transmitter} shows the design of the transmitter side. The \textit{server side} is on the top left corner, where the entire scenario is established, including how many terminals there are in the emulation, the roles of the terminals (transmitter, ground hub receiver or jammer), the link's properties (carrier frequency, bandwidth and SNR level ) and other situation parameters. The server side is written and compiled in java to support the fast processing speed. 

The next part on the top right side is the \textit{PHY signal processing}. It generates the transmitting signals in binary format, and sends it through channel coding, framing and modulation. In addition, the PHY layers convert the scenario set up into parameter settings in order to pass to next layer. This component is written in Matlab. The reason we choose to use Matlab is that we have developed many coding and modulation tools in Matlab\cite{Xiong2016, Xiong20162}. It is convenient to utilize these tools to implement reducing the developing time and risks. The connection between server and the PHY signal processing is the high speed DDS which serves the communication between components written with any language. We successfully established the connection between java and Matlab, and the emulation design could be extended to other languages for cross platform development. 

The bottom right part of Figure \ref{fig:transmitter} is the RF front end controlling unit. This component carries out two tasks: (1) taking the configuration setting from PHY signal processing and deploying the parameters like amplifying gain, carrier frequencies, and sampling rates; and (2) taking data input from the PHY signal processing and performing data preparation for transmitting. Since this part is directly connected with the USRP and the antenna, we choose to use the GNU radio to establish the RF controller module. The  module is built with a combination of C++ and python. C++ contributes to the signal processing modules in the module, and python does the wrapping. The connection between the PHY signal processing part and the GNU radio is through the database. We choose to use the database as an interface because: (1) The connection is cross platform between components written by different languages, (Matlab, C++ and python); (2) the information exchange in the connection has various formats.  These formats include the data which needs to be transmitted in binary format, the bandwidth and frequencies are integers, and the USRP device serial number is a string. We build multiple tables in the database in order to accommodate all these requirements. The modules in the GNU radio takes the information they need from the designated table without interfering other modules.

The final part is the RF front end controlled by the GNU radio. It performs the radio waveform transmission, as shown in the bottom left of Figure \ref{fig:transmitter}.

It can be seen that the connection between any two modules shown in Figure \ref{fig:transmitter} is defined in a way that any component module can be connected. In other words, it is possible to replace any part with an other form if we want to develop a different emulation testbed. The modules are loosely connected through either DDS or the database, but the latency and  communication speed are not sacrificed. Any part of the design reacts to the changes in real time. Also in the GNU radio, we develop modules individually, and interconnected them with python. The modules can be modified or replaced to fit any condition and requirement.

Another advantage of the design is top to bottom control mechanisms. The scenario and emulation status is defined and visualized in the server. It initiates the emulation, sends out commands, and monitors the whole process. The rest of the testbed follows the instructions and updates the real time status of the emulation.

\begin{figure}[t]
  \centering
    \includegraphics[width=0.8\linewidth]{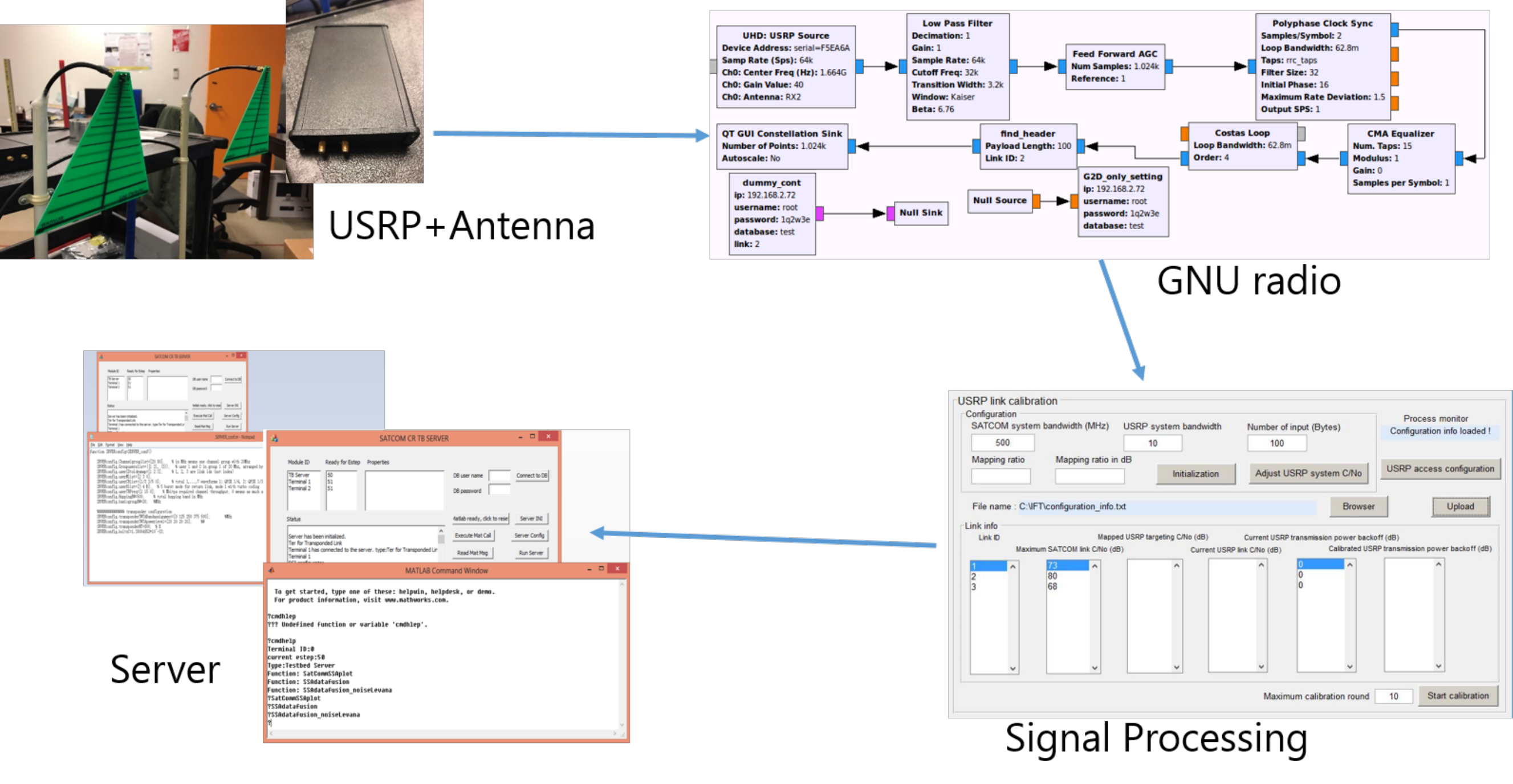}
    \caption{Receiver}
    \label{fig:receiver}
\end{figure}

Figure \ref{fig:receiver} shows the receiver side structure. It is similar to the transmitter side. However, the modules that are placed in each part are different. For example, in the GNU radio, frequency and frame synchronization is included to recover signal and locate the payload.

\section{Link Calibration}
\label{sec:calibration}
In order to emulate the correct scenario, the systems needs to establish each link with desired signal-to-noise (SNR) ratio. In  realistic transceiving, the noise floor is not controllable. Also it is not efficient to measure the pathloss each time before a test. Instead, we first estimate the current link SNR with some initial power level, adjust the transmitting power if the current SNR is different from desired value. We use the emulation modules and structure described  above to fulfill the calibration.

\begin{figure}[t]
  \centering
    \includegraphics[width=0.8\linewidth]{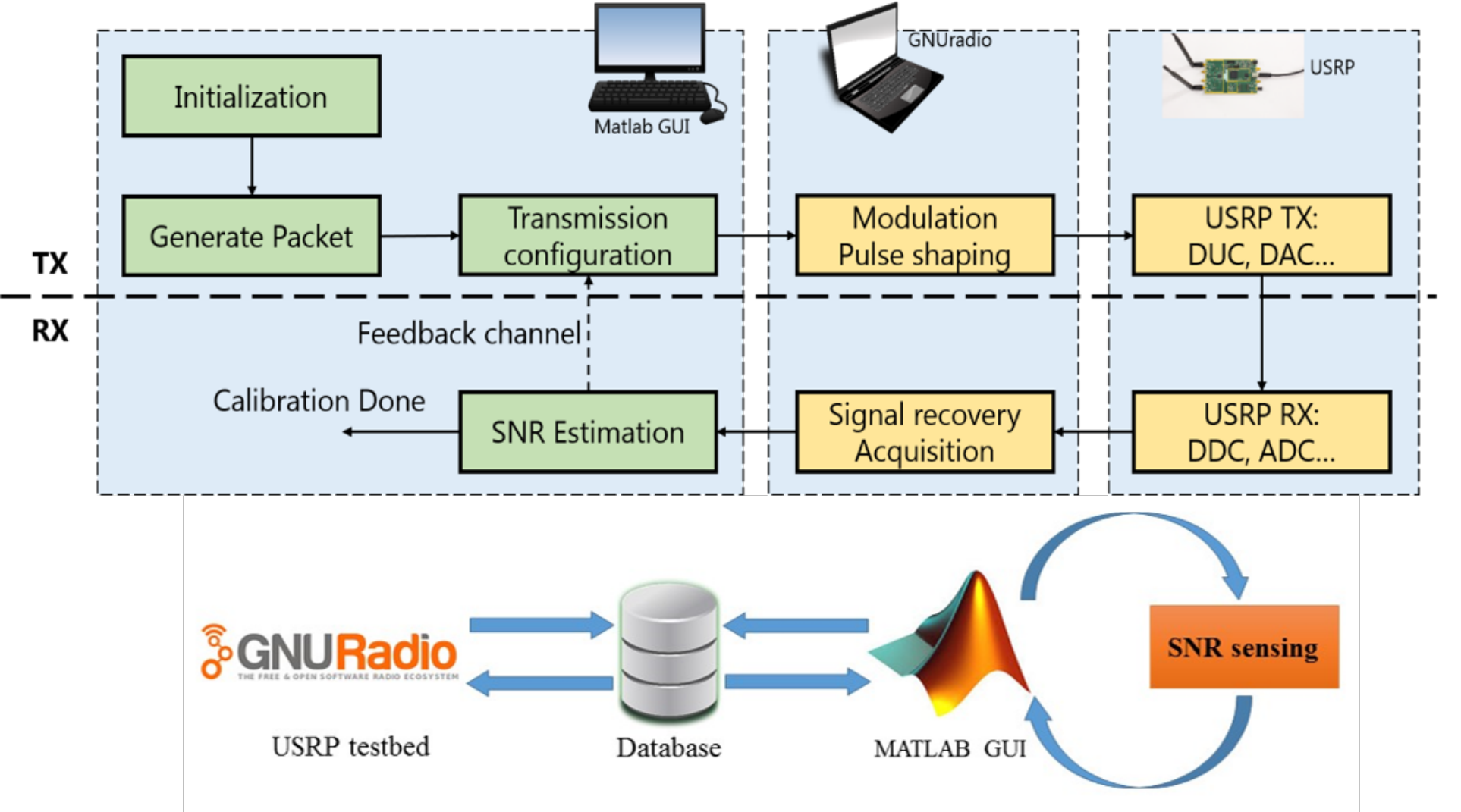}
    \caption{Link Calibration}
    \label{fig:calibration}
\end{figure}

As shown in Figure \ref{fig:calibration}, the TX (transmitter) and RX (receiver) work together for system calibration. Initialization takes the scenario information and configures the emulation. The rest of the transceiving follows a similar route until the signal is recovered at the receiver. The SNR level of the current channel is estimated and sent back to the transmitter using the feedback channel. In this way, the transmitter knows if the current power is too high or too low compared to the defined scenario. The loop keeps running until the link SNR reaches the requirement. 

\begin{figure}[t]
  \centering
    \includegraphics[width=0.7\linewidth]{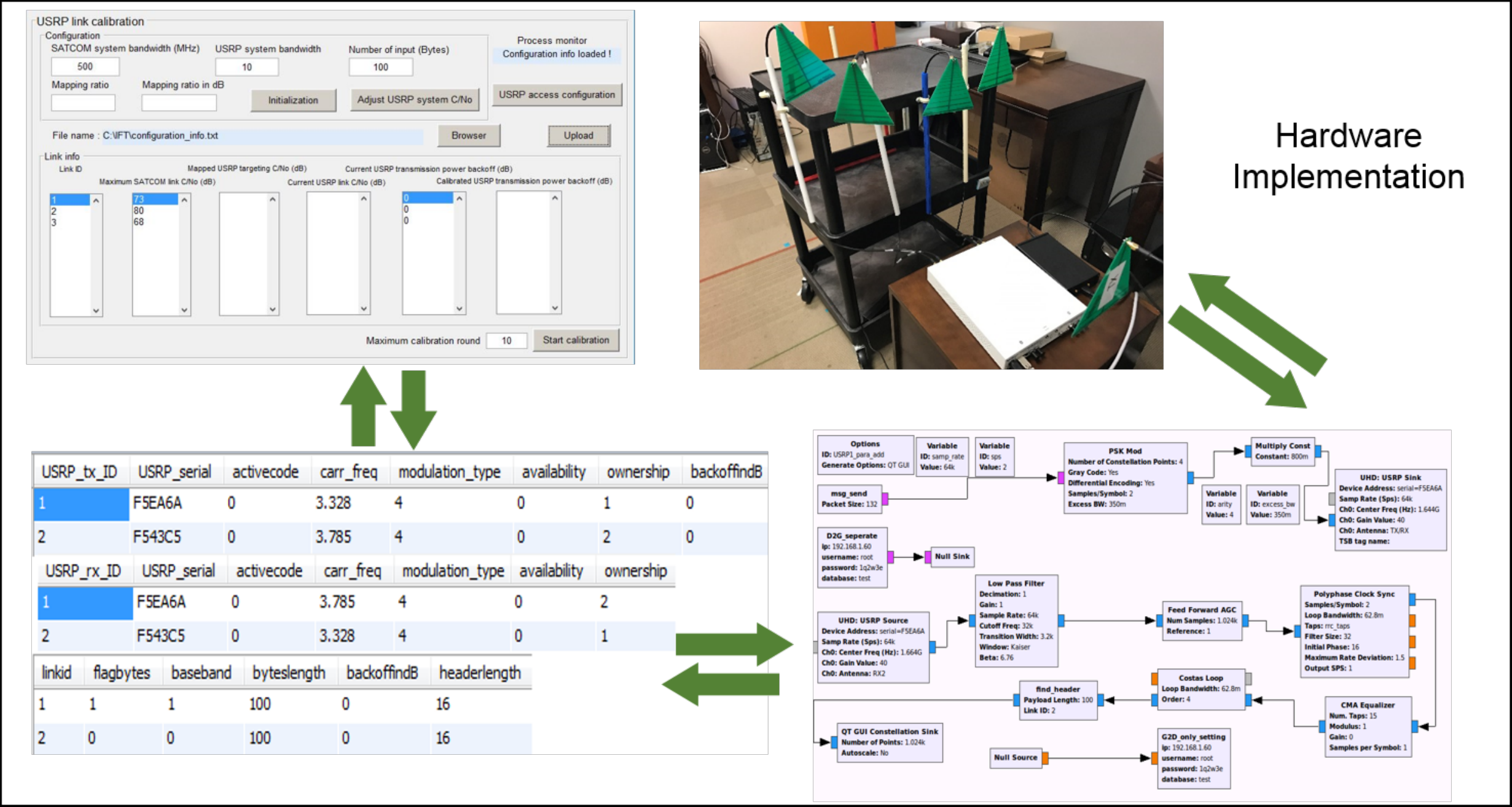}
    \caption{PHY processing and GNU radio}
    \label{fig:calibration2}
\end{figure}

Figure \ref{fig:calibration2} shows the detailed procedure between PHY processing and GNU radio of the calibration. The screen shot on the left bottom corner establishes the database for the configuration information where the transmitter and receiver information is stored in different tables in order to avoid confusion. The row of the table is identified by its identifier (ID) and USRP serial number which are unique. Other information like carrier frequency and modulation type are also defined. The ownership section in the database is created to represent the assignment of USRP to particular link. For example, link 1 is the uplink from terminal to transponder. If USRP 1 is assigned to link on in transmitter table, it will perform transmitting for terminal uplink with corresponding configurations. Here the link could be a jamming link or transmitting link. The difference is how the link is configured.

\section{Conclusions}
\label{sec:conclusion}
In this paper, we proposed a standard interface for the SATCOM software defined radio (SDR) emulation testbed. We defined the testbed in three layers due to different functions that are performed. The connection between layers are designed to accommodate cross platform and multi-language development. We utilized the high speed data distributed service (DDS) and database to interact between layers and achieved multiple format information exchange. We also designed the top to bottom structure for ease of controlling and monitoring. To show the feasibility and performance of the design, we demonstrated a link calibration using the structure and interfaces proposed in this article.

\textbf{Acknowledgments}: The work was supported under contract FA9453-14-C-0017. The views and conclusions contained herein are those of the authors and should not be interpreted as necessarily representing the official policies, either expressed or implied, of AFRL, or the U.S. Government. 

% References

%\bibliography{report} % bibliography data in report.bib

\begin{thebibliography}{1}

\bibitem{Tian2012}
Tian, Z., and Blasch, E., "Compressed Sensing for MIMO Radar: A Stochastic Perspective," \emph{IEEE Statistical Signal Processing Workshop (SSP)}., (2012).


\bibitem{Chenji2016}
Chenji, H., Stewart, G., Wu, Z., Javaid, A., Devabhaktuni, V., and Kul, B., "An Architecture Concept for Cognitive Space Communication Networks," \emph{AIAA International Communications Satellite Systems Conference}., Oct, (2016).

\bibitem{Gardellin2012}
Gardellin, V., Mammasis, K., Martelli, F., and Santi, P., "The MIMONet Software Defined Radio Testbed," \emph{Proc. InfQ}., (2012).


\bibitem{Baruffa2014}
Baruffa, G., Rugini, L., and Banelli, P., "Design and Validation of a Software Defined Radio Testbed for DVB-T Transmission," \emph{Radioengineering}., vol. 23, No. 1, April. (2014).

\bibitem{Weiss2003}
Weiss, S., Shligersky, A.,  Abendroth, S., Reeve, J., Moreau, L., Dodgson, T., E., and Babb, D., "A Software Defined Radio Testbed Implementation," \emph{IEE Colloquium on DSP enabled Radio}., Sept (2003).

\bibitem{Shen2014}
Shen, D., Chen, G.,  Wang, G., Pham, K., Blasch, E., and Tian, Z., "Network Survivability Oriented Markov Games (NSOMG) in Wideband Satellite Communications," \emph{IEEE/AIAA Digital Avionics Systems Conference}., (2014).

\bibitem{Liu20152}
Lu, J., and Niu, R., "A State Estimation and Malicious Attack Game in Multi-Sensor Dynamic Systems," \emph{IEEE Information Fusion}., July (2015).

\bibitem{Han2016}
Han, T., and Ansari, N., "A Traffic Load Balancing Framework for Software-Defined Radio Access Networks Powered by Hybrid Energy Sources," \emph{IEEE/ACM Transactions on Networking}.,  Vol. 24, Issue 2,
pp. 1038-1051, (2016).

\bibitem{Liu2015}
Lu, J., and Niu, R., "False information detection with minimum mean squared errors for Bayesian estimation," \emph{49th Annual Conference on Information Sciences and Systems (CISS)}., March (2015).

\bibitem{Ceylan2016}
Ceylan, O., Gannon, A., et al, "Small Satellites Rock A Software-Defined Radio Modem and Ground Station Design for Cube Satellite Communication," \emph{IEEE Microwave Magazine}.,  Vol. 17, Issue 3,
pp. 26-33, (2016).

\bibitem{Liu2014}
Lu, J., and Niu, R., "False information injection attack on dynamic state estimation in multi-sensor systems," \emph{IEEE Information Fusion}., July (2014).

\bibitem{Steward2015}
Steward, R., Crockett, L., et al, "A low-cost desktop software defined radio design environment using MATLAB, simulink, and the RTL-SDR," \emph{IEEE Communications Magazine}.,  Vol. 53, Issue 9,
pp. 64-71, (2015).

\bibitem{Liu20153}
Lu, J., and Niu, R., "Malicious attacks on state estimation in multi-sensor dynamic systems," \emph{IEEE International Workshop on Information Forensics and Security}., Dec (2015).

\bibitem{Macedo2015}
Macedo, D., Guedes, D., et al, "Programmable Networks—From Software-Defined Radio to Software-Defined Networking," \emph{IEEE Communications Surveys and Tutorials}., Vol. 17, Issue 2, pp. 1102-1125, (2015).

\bibitem{Xia2015}
Xia, S., and Wang, P., "Distributed throughput optimal scheduling in the presence of heavy-tailed traffic," \emph{IEEE International Conference on Communication}., (2015).

\bibitem{Liu20162}
Lu, J., and Niu, R., "Sparse attacking strategies in multi-sensor dynamic systems maximizing state estimation errors," \emph{IEEE International Conference on Acoustics, Speech and Signal Processing }., March (2016).

\bibitem{Zhang2015}
Zhang, J., Yang, L., Hanzo, L., and Gharavi, H., "Advances in Cooperative Single-Carrier FDMA Communications: Beyond LTE-Advanced," \emph{IEEE Communications Surveys and Tutorials}.,  vol. 17, no. 2, pp. 730-756, Secondquarter (2015).

\bibitem{Drozdenko2015}
Drozdenko, B., Subramanian, R., Chowdhury, K., and Leeser, M., "Implementing a MATLAB-based Self-Configurable
Software Defined Radio Transceiver," \emph{International Conference on Cognitive Radio Oriented Wireless Networks}., Oct. (2015).

\bibitem{Wei2016}
Wei, X., Liu, H., Geng, Z., et al, "Software Defined Radio Implementation of a Non-Orthogonal Multiple Access System Towards 5G," \emph{IEEE Access}.,  Vol. 4,
pp. 9604-9613, Dec. (2016).

\bibitem{Wei2007}
Wei, M., Chen, G., Cruz, J., et al, "Multi-Pursuer Multi-Evader Pursuit-Evasion Games with Jamming Confrontation," \emph{AIAA Journal of Aerospace Computing, Information, and Communication}.,  Vol. 4, No. 3, pp. 693 – 706, (2007).

\bibitem{Tian20122}
Tian, X., Tian, Z., Pham, K., et al, "Jamming/Anti-jamming Game with a Cognitive Jammer in Space Communication," \emph{Proc. SPIE}.,  Vol. 8385 (2012).

\bibitem{Yu2013}
Yu, W., Fu, X., Blasch, E., et al, "On Effectiveness of Hopping-Based Techniques for Network Forensic Traceback," \emph{International Journal of Networked and Distributed Computing}.,  Vol. 1, No. 3, (2013).

\bibitem{Shen2012}
Shen, D., Chen, G., Pham, K., and Blasch, E., "Models in frequency-hopping-based proactive jamming mitigation in space communication networks," \emph{Proc. SPIE}.,  Vol. 8385 (2012).

\bibitem{Zhang2013}
Zhang, J., Yang, L., and Hanzo, L., "Energy-Efficient Dynamic Resource Allocation for Opportunistic-Relaying-Assisted SC-FDMA Using Turbo-Equalizer-Aided Soft Decode-and-Forward," \emph{IEEE Transactions on Vehicular Technology}.,  vol. 62, no. 1, pp. 235-246, Jan. (2013).

\bibitem{Xiong2016}
Xiong, W., Mo, Z., Chen, G., et al, "Agile MU-MIMO in congested environments with robust channel estimation," \emph{IEEE MILCOM}, (2016).

\bibitem{Xiong20162}
Xiong, W., Wang, G., Tian, X., et al, "Hybrid onboard and ground based digital channelizer beam-forming for SATCOM interference mitigation and protection," \emph{Proc. SPIE}.,  vol. 9838. (2016).















\end{thebibliography}
%\bibliographystyle{spiebib} % makes bibtex use spiebib.bst

\end{document}